\documentclass[runningheads]{llncs}
\usepackage[T1]{fontenc}
\usepackage{graphicx}
\usepackage{booktabs}
\usepackage{tabularx}
\usepackage{amsmath}
\usepackage{multirow}
\usepackage{diagbox}
\usepackage{upgreek}
\usepackage[hidelinks]{hyperref}
\begin{document}
\title{MIP-Based Tumor Segmentation: A Radiologist-Inspired Approach}

\author{
Romario Zarik\inst{1}\orcidID{0009-0009-5488-4538} \and
Nahum Kiryati\inst{2}\orcidID{0000-0003-1436-2275} \and
Michael Green\inst{3}\orcidID{0000-0003-3919-970X} \and
Liran Domachevsky\inst{3} \and
Arnaldo Mayer\inst{3}\orcidID{0000-0003-3702-1764}
}

\authorrunning{Zarik et al.}

\institute{
School of Electrical and Computer Engineering, Tel Aviv University, Tel Aviv, Israel\\
\email{romariozarik@mail.tau.ac.il}
\and
Klachky Chair of Image Processing, School of Electrical and Computer Engineering, Tel Aviv University, Tel Aviv, Israel\\
\email{nk@eng.tau.ac.il}
\and
Diagnostic Imaging, Sheba Medical Center, affiliated with the Gray School of Medicine, Tel Aviv University, Tel Aviv, Israel\\
\email{liran.domachevsky@sheba.health.gov.il, arnaldo.mayer@sheba.health.gov.il}
}

\maketitle              
\begin{abstract}
PET/CT imaging is the gold standard for tumor detection, offering high accuracy in identifying local and metastatic lesions. Radiologists often begin assessment with rotational Multi-Angle Maximum Intensity Projections (MIPs) from PET, confirming findings with volumetric slices. This workflow is time-consuming, especially in metastatic cases. Despite their clinical utility, MIPs are underutilized in automated tumor segmentation, where 3D volumetric data remains the norm. We propose an alternative approach that trains segmentation models directly on MIPs, bypassing the need to segment 3D volumes and then project. This better aligns the model with its target domain and yields substantial gains in computational efficiency and training time. We also introduce a novel occlusion correction method that restores MIP annotations occluded by high-intensity structures, improving segmentation. Using the autoPET 2022 Grand Challenge dataset, we evaluate our method against standard 3D pipelines in terms of performance and training/computation efficiency for segmentation and classification, and analyze how MIP count affects segmentation. Our MIP-based approach achieves segmentation performance on par with 3D ($\leq$1\% Dice difference, 26.7\% better Hausdorff Distance), while reducing training time (convergence time) by 55.8–75.8\%, energy per epoch by 71.7–76\%, and TFLOPs by two orders of magnitude, highlighting its scalability for clinical use. For classification, using 16 MIPs only as input, we surpass 3D performance while reducing training time by over 10× and energy consumption per epoch by 93.35\%. Our analysis of the impact of MIP count on segmentation identified 48 views as optimal, offering the best trade-off between performance and efficiency.

\end{abstract}

\section{Introduction}
\label{sec:intro}

Positron emission tomography (PET), combined with CT or MRI, has become an invaluable tool in cancer imaging, allowing the assessment of primary tumors and metastatic disease across the whole body. In PET imaging, a radiotracer consisting of a positron-emitting isotope bound to an organic ligand, such as Fluorodeoxyglucose (FDG), is intravenously injected. The photon pairs resulting from positron annihilation are recorded by a ring of detectors surrounding the body. Through tomographic reconstruction, a 3D distribution map of the tracer’s physiological uptake is generated \cite{crișan2022radiopharmaceuticals}.

\begin{figure}[t]
    \begin{minipage}[b]{1.0\linewidth}
     \centering 
     {}
     \centerline{\includegraphics[width=8.5cm]{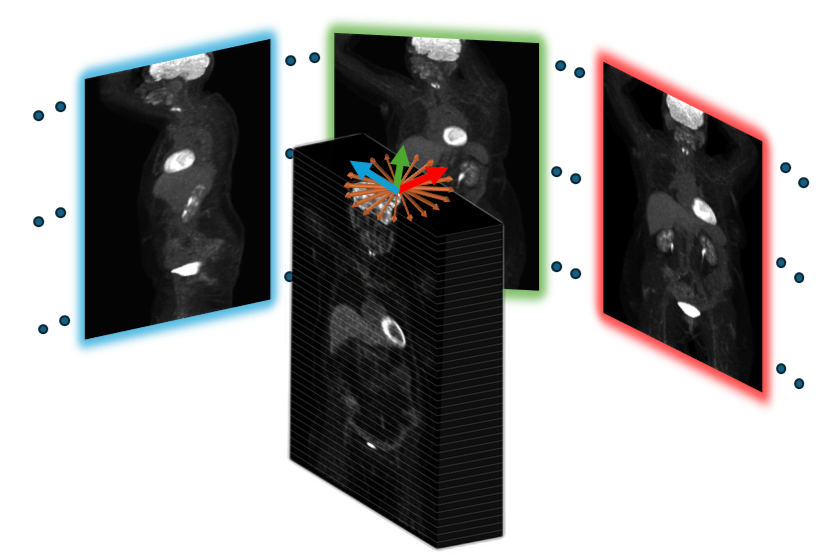}} 
     \caption{Illustration of Multi-Angle Maximum Intensity Projections (MIPs) derived from the same 3D PET volume. Each MIP corresponds to a unique yaw angle, capturing different perspectives of the tumor. The figure presents three example MIPs generated at 0° (red), 45° (green), and 90° (cyan).}
      \label{fig:MIPs}
    \end{minipage}
\end{figure}

To facilitate clinical assessment of a patient's PET scan, radiologists frequently use Multi-Angle Maximum Intensity Projections (MIPs), which are 2D images that capture the highest intensity values along each line of sight, making it easier to identify regions with increased tracer uptake. This technique enhances the visualization of structures with high radiotracer uptake, such as tumors, by providing a comprehensive overview of their distribution and intensity. By computing MIPs for different yaw angles, we obtain a series of 2D MIP images (see Fig.\!~\ref{fig:MIPs}). Even before consulting patient history or indication, the initial review of a continuously rotating MIP is valuable in achieving an unbiased assessment, favoring a “gestalt” perception of the case \cite{hofman2016we}. 

Interpreting PET scans is a time-consuming process that often requires cross-referencing with CT or MRI. Automatically delineating potential tumors directly on the rotating MIPs can streamline this workflow by guiding radiologists to regions suspected of containing active lesions, thereby significantly reducing medical interpretation time.

Although deep learning has been applied to PET-CT for lesion detection and disease staging, most models treat PET as a slice-based modality, overlooking the benefits of rotated MIP images. A few recent studies have incorporated MIPs into their pipelines. For instance, Kawakami et al.\ \cite{kawakami2020development} used a YOLOv2 detector on multi-orientation MIPs to identify physiological uptake and highlight suspect regions. Takahashi et al.\ \cite{takahashi2022deep} employed an Xception model on MIPs from multiple angles to improve breast cancer classification. In another study, Heiliger et al.\ \cite{heiliger2022autopet} combined sagittal and coronal MIPs with a gating strategy to reduce false positives. Toosi et al.\ \cite{toosi2024fully} extracted features from 72 MIPs and used them in a recurrence-free survival prediction pipeline for head and neck cancer.

While 3D volumetric segmentation remains the dominant paradigm, recent studies have begun exploring 2D projections to leverage their efficiency and complementary information. Constantino et al. \cite{constantino2025use} demonstrated that incorporating MIP representations can enhance 3D PET/CT segmentation for both FDG and PSMA tracers, and Toosi et al. \cite{toosi2024segment} reconstructed volumetric segmentations from 2D MIPs, achieving state‑of‑the‑art results on PSMA PET scans. Similarly, Wang et al. \cite{wang20232} introduced a “2.75D” approach, combining multiple 2D views to enrich 3D learning in data‑limited scenarios. Building on these insights, our study focuses on direct segmentation of MIPs: we systematically analyze their standalone performance, evaluate how the number of MIPs impacts segmentation quality, and benchmark their efficiency against traditional 3D pipelines on FDG PET scans, which are widely used across cancer types.

Training directly on Multi-Angle Maximum Intensity Projections (MIPs) offers substantial advantages over conventional 3D volumetric segmentation. MIPs provide a computationally efficient representation of PET data; for instance, 16 views capture approximately 4\% of the volumetric information (depending on resolution) while preserving critical lesion details. This compact format enables training on standard hardware with significantly lower memory and computational demands. Moreover, operating on 2D projections reduces training time and facilitates faster model development and deployment. The resulting lightweight models are more suitable for real-time clinical use. Importantly, aligning the model’s input domain with radiologists’ diagnostic perspective enhances the learning of clinically relevant features.

The major contributions of this work are as follows:
\begin{enumerate}
    \item \textbf{Segmentation Performance Comparison:} We demonstrate that direct segmentation on MIPs (48 MIPs) achieves performance comparable to conventional 3D segmentation followed by MIP projection, while significantly reducing training time, energy consumption, and computational cost.
    
    \item \textbf{Classification Performance Comparison:} To further highlight the effectiveness of MIPs, we show that a CNN trained on just 16 MIPs outperforms a 3D model in binary classification (healthy vs.\ non-healthy), achieving over 10× faster training and a significant reduction in energy use per epoch.

    \item \textbf{Occlusion Correction for MIP Annotations:} We introduce a pre-processing method that corrects MIP annotations occluded by high-intensity structures, reducing false positives and improving segmentation accuracy.

    \item \textbf{Impact of MIP Quantity and Angular Resolution:} We analyze the effect of varying the number of MIPs, equivalent to adjusting angular resolution or projection spacing, and identify 48 views as the optimal trade-off between segmentation performance and efficiency.
\end{enumerate}

\noindent The remainder of this paper is organized as follows: Section 2 details our methodology, Section 3 describes the experimental setup and results, and Section 4 discusses conclusions, limitations, and future research directions.

\section{Methods}
\label{sec:format}

\subsection{Data Partitioning} 

\noindent \textbf{Data.} We used the open-source autoPET 2022 dataset \cite{gatidis2022whole}, which includes 1,014 PET/CT scans from 900 patients diagnosed with lung cancer, lymphoma, melanoma, or confirmed as healthy (Table \ref{tab:dataset_distribution}). Each case provides a 3D PET scan, CT scan, and tumor segmentation map. While MIPs can be generated for both PET and CT, we focused on PET MIPs, as they are standard for whole-body tumor assessment and enable a fair comparison with 3D PET-based segmentation. PET values were standardized to SUV units \cite{fletcher2010pet}.

\noindent \textbf{Data Split.} We partitioned the dataset by setting aside 15\% of the data as an independent test set for the final evaluation. The remaining 85\% were used for 5-fold cross-validation, ensuring that each model was trained and validated on different subsets before being tested on the independent test set. To maintain consistency, we preserved the same class distribution across all splits.

\begin{table}[h]
    \centering
    \caption{Distribution of cases across the four classes in the dataset}
    \label{tab:dataset_distribution}
    \begin{tabularx}{\textwidth}{l>{\centering\arraybackslash}X>{\centering\arraybackslash}X>{\centering\arraybackslash}X>{\centering\arraybackslash}X}
        \toprule
        \textbf{Class} & \textbf{\begin{tabular}[c]{@{}c@{}}Negative\\(Healthy)\end{tabular}} & \textbf{Lymphoma} & \textbf{Melanoma} & \textbf{\begin{tabular}[c]{@{}c@{}}Lung\\Cancer\end{tabular}} \\
        \midrule
        \textbf{Number of Cases} & 513 & 145 & 188 & 168 \\
        \bottomrule
    \end{tabularx}
\end{table}

\subsection{MIP generation}

\noindent To generate MIPs for both PET images and their lesion annotations, we applied maximum-intensity projections along the anterior-posterior axis of the 3D data after rotating it around the superior-inferior (vertical/yaw) axis. Each MIP image, \( F_{k}(i,j) \), is computed by taking the maximum intensity across the depth dimension \( d \) for each pixel \((i, j)\) and MIP index \( k \), as defined in Eq.~\ref{eq:mip_computation}:

\begin{equation}
F_{k}(i,j) = \max_{d} f_{k}(i, j, d)
\label{eq:mip_computation}
\end{equation}

\noindent where \( F_{k}(i,j) \) represents the resulting MIP image for index \( k \), and \( f_{k}(i, j, d) \) denotes the 3D data after rotation by an angle of $k \Delta \Theta$. The MIP images were captured at equal angular increments, $\Delta \Theta$, spanning from 0 to 180 degrees to ensure symmetry. The angular step size, $\Delta \Theta$, is given by Eq.~\ref{eq:angular_increments}:

\begin{equation}
\Delta \Theta (N) =  \frac{180^{\circ}}{N}
\label{eq:angular_increments}
\end{equation}

\noindent where $N$ is the number of MIPs created from the 3D data. Images from 180 to 360 degrees were not captured, as they are mirror images of those in the 0–180 degree range. In our study, we generated MIP image sets in multiples of 16, ranging from 16 to 80 images, all derived from the 0–180 degree rotation range.

\subsection{MIP occlusion correction}

Projecting 3D annotations onto MIPs sometimes led to inconsistent labels due to occlusions by high-intensity structures. These occlusions occur when lesions and high-FDG-uptake organs (e.g., brain, heart, and kidneys) are aligned along the same viewing angle used to generate the MIP. As a result, high-intensity pixels from these organs can dominate the projection, obscuring tumor annotations and introducing false positives. To address this, we designed a three-step processing pipeline to generate a more accurate training set:

\begin{enumerate}
    \item \textbf{Occlusion Detection}: For each labeled tumor (connected component) in the generated MIPs, we verified that at least 75\% of the pixels labeled as tumor originated from tumors in the volumetric PET data. This verification was possible because the projection process allows us to trace each pixel back to its corresponding voxel in the 3D data.
    \item \textbf{Annotation Splitting}: If a labeled tumor contained fewer than 75\% tumor-originated pixels, we split the label to retain only the pixels confirmed to originate from the tumor.
    \item \textbf{Low-Contrast Filtering}: Additional processing was applied to handle remnants of tumors with very low contrast, which were deemed undetectable by eye and subsequently removed from the annotations.
\end{enumerate}

\noindent This process ensured that only accurate and visible tumor annotations were retained in the MIP images, as occluded tumors are generally considered undetectable by radiologists. Analysis of the annotations revealed that only 0.57\% of tumors were excluded from all MIPs during this preprocessing step, accounting for just 0.09\% of the original volumetric volume. These figures suggest that the GT was largely preserved. In our experiments, we utilize both the original GT MIPs (OR-MIPs) and the occlusion-corrected MIPs (OC-MIPs) during training.

\subsection{Models \& Implementation} 
\textbf{Models.} For semantic segmentation on MIPs, we employed Convolutional Neural Networks (CNNs) due to their flexibility and robustness across different numbers of MIPs. Among the CNN architectures evaluated, Attention U-Net \cite{oktay2018attention} demonstrated the best performance and was therefore selected for all MIP-based segmentation experiments. For 3D volumetric segmentation, we implemented the training pipeline that ranked 5th in the autoPET 2022 Grand Challenge \cite{heiliger2022autopet}, which was based on a Swin-UNETR \cite{hatamizadeh2021swin} architecture. Higher-ranked teams either did not share their code for reproducibility or relied on high-VRAM GPUs, which were not available to us. For classification, we used a CNN encoder followed by attention pooling to produce a fixed-size representation regardless of input shape. This was passed through a fully connected head ending with two output neurons for binary prediction. 

\noindent \textbf{Implementation.} The implementation used the MONAI \cite{cardoso2022monai} and PyTorch \cite{paszke2019pytorch} frameworks, with additional support from the original GitHub repositories of the models. All models were trained using a single 24GB RTX 3090 GPU. The code and preprocessing tools used in this study are publicly available in our \href{https://github.com/ZROM-GIT/MIP-Based-Tumor-Segmentation}{GitHub repository}\footnote{\url{https://github.com/ZROM-GIT/MIP-Based-Tumor-Segmentation}}.

\subsection{Loss Function \& Evaluation}
\textbf{Loss Function.} Segmentation models were trained using Dice Loss~\cite{milletari2016v}, either alone or in combination with Cross-Entropy Loss~\cite{long2015fully}, depending on the experimental setting. For classification, we used Cross-Entropy Loss exclusively.

\noindent \textbf{Evaluation.} We report Dice Score \cite{sorensen1948method}, IoU \cite{jaccard1901etude}, and Hausdorff Distance (HD) \cite{huttenlocher1993comparing} for segmentation; Accuracy, Precision, Recall \cite{sokolova2009systematic}, and F1-score \cite{van1979information} for classification; and convergence time (CT), time per epoch (TPE), and energy per epoch (EPE) for training efficiency. Convergence time is defined as the time taken to reach peak validation performance (Dice/Accuracy), with the corresponding weights used for final testing. Inference time and computational cost were measured using the time and ptflops \cite{ptflops} python packages, respectively. Costs are reported in teraflops (TFLOPs), where 1 TFLOP equals 10¹² floating-point operations. MIP generation costs were included for fair evaluation.

\section{Experiments and Results}

\subsection{MIPs vs.\ 3D}
We compare lesion segmentation trained on 3D volumetric data (followed by maximum-intensity projection) to direct segmentation on MIPs. The 3D pipeline from \cite{heiliger2022autopet} was adapted to use only the PET (SUV) channel, and its predictions were projected to MIPs for a shared 2-D evaluation domain.

\begin{table}[ht]
\centering
\caption{Comparison results (mean ± standard deviation) of 3D-based and MIP-based models' segmentation performance and efficiency metrics. All trained models were tested on the OR-MIPs test set for the fairest comparison.}
\label{tab:mips_vs_3d_segmentation_results}
\renewcommand{\arraystretch}{1.2}
\setlength{\tabcolsep}{6pt}
\begin{tabular}{p{0.2cm}lccc}
\toprule
\textbf{} & \textbf{Metric} & \textbf{3D projected} & \textbf{OR-MIPs} & \textbf{OC-MIPs} \\
\midrule
\multirow{3}{*}{\rotatebox{90}{Segment.}}
    & Dice (↑)                   & \textbf{0.597} ± 0.05 &  0.578 ± 0.01 & 0.591 ± 0.01 \\
    & IoU (↑)                    & \textbf{0.471} ± 0.04 & 0.452 ± 0.01 & 0.466 ± 0.01 \\
    & HD (↓)     & 139.614 ± 8.42 & 102.813 ± 9.61 & \textbf{102.26} ± 9.53 \\
\midrule
\multirow{3}{*}{\rotatebox{90}{Efficiency}}
    & CT (hours, $\downarrow$)       & 54.64 ± 19.22 & 24.14 ± 17.8 & \textbf{13.18} ± 4.1 \\
    & EPE (\(\frac{\text{Wh}}{\text{epoch}}\), $\downarrow$)       & 142.2 ± 79.1 & 40.22 ± 12.48 & \textbf{34.194} ± 4.7 \\
    & TFLOPs ($\downarrow$)       & 317.42 ± 144.05 & \textbf{0.97} ± 0.29 & \textbf{0.97} ± 0.29 \\

\bottomrule
\end{tabular}
\end{table}

All models were evaluated on the OR-MIPs dataset, which offers the fairest basis for comparison: once 3D predictions are projected, occlusions cannot be corrected as in OC-MIPs, so evaluating in a shared domain avoids bias against the 3D model. Results (Table~\ref{tab:mips_vs_3d_segmentation_results}) show comparable Dice scores across methods, with mean values within 1\% of each other. Both MIP-based models achieved better Hausdorff distances, indicating improved boundary accuracy. A one-sided Wilcoxon signed-rank test on fold-wise paired results (null: no difference; alternative: 3D is better) yielded p-values of 0.22 (OR-MIPs) and 0.5 (OC-MIPs), showing no statistical advantage for 3D.

\noindent In terms of efficiency, the OC-MIPs model reduced training time from 54.6 to 13.2 hours (4.1× faster), energy per epoch from 142.2 to 34.2 Wh (4.2× lower), and TFLOPs from 317.4 to 0.97 (over 300× lower), demonstrating substantial gains in scalability and resource savings.

To further highlight the effectiveness of MIPs, we compare classification performance using identical CNN-based models with attention pooling and a FC head, trained on either 3D volumes or 16-angle MIPs (healthy vs.\ non-healthy). As shown in Table \ref{tab:mips_vs_3d_classification_results}, MIP-based models outperform their 3D counterparts across most metrics, including accuracy (80.5\% vs.\ 72.8\%) and F1-score (86.4\% vs.\ 82.3\%). While the 3D models show slightly higher recall (91.9\% vs.\ 89.5\%), their large standard deviation suggests instability and less consistent performance. In terms of classification training efficiency, MIP-based training reduced training time by over 10× and energy consumption per epoch by 93.35\%.

\begin{table}[ht]
\centering
\caption{Classification results (mean ± standard deviation) and training efficiency metrics of identical models trained on the 3D and 16 MIPs datasets.}
\label{tab:mips_vs_3d_classification_results}
\setlength{\tabcolsep}{6pt} 
\renewcommand{\arraystretch}{1.2} 
\begin{tabular}{p{0.2cm}llcc} 
\toprule
\textbf{} & \textbf{Metric} & \textbf{3D Dataset} & \textbf{16 MIPs Dataset} \\
\midrule
\multirow{4}{*}{\centering\rotatebox{90}{Classification}} 
    & Accuracy (\%, $\uparrow$)         & 72.8 ± 3.2 & \textbf{80.5} ± 1.7 \\
    & Precision (\%, $\uparrow$)        & 75.4 ± 6.0 & \textbf{83.6} ± 3.3 \\
    & Recall (\%, $\uparrow$)           & \textbf{91.9} ± 8.8 & 89.5 ± 2.9 \\
    & F1-score (\%, $\uparrow$)         & 82.3 ± 1.2 & \textbf{86.4} ± 0.8 \\
\midrule
\multirow{3}{*}{\centering\rotatebox{90}{Efficiency}} 
    & CT (hours, $\downarrow$)     & 44.7 ± 1.5 & \textbf{4.2} ± 0.2 \\
    & TPE (\(\frac{\text{minutes}}{\text{epoch}}\), $\downarrow$)     & 26.1 ± 0.59 & \textbf{1.45} ± 0.04 \\
    & EPE (\(\frac{\text{Wh}}{\text{epoch}}\), $\downarrow$)     & 70.7 ± 3.23 & \textbf{4.7} ± 0.1 \\
\bottomrule
\end{tabular}
\end{table}


\subsection{Number of MIPs}

\label{sec:nmips}
\begin{figure}[htb]
    \centering
    \includegraphics[width=6.0cm]{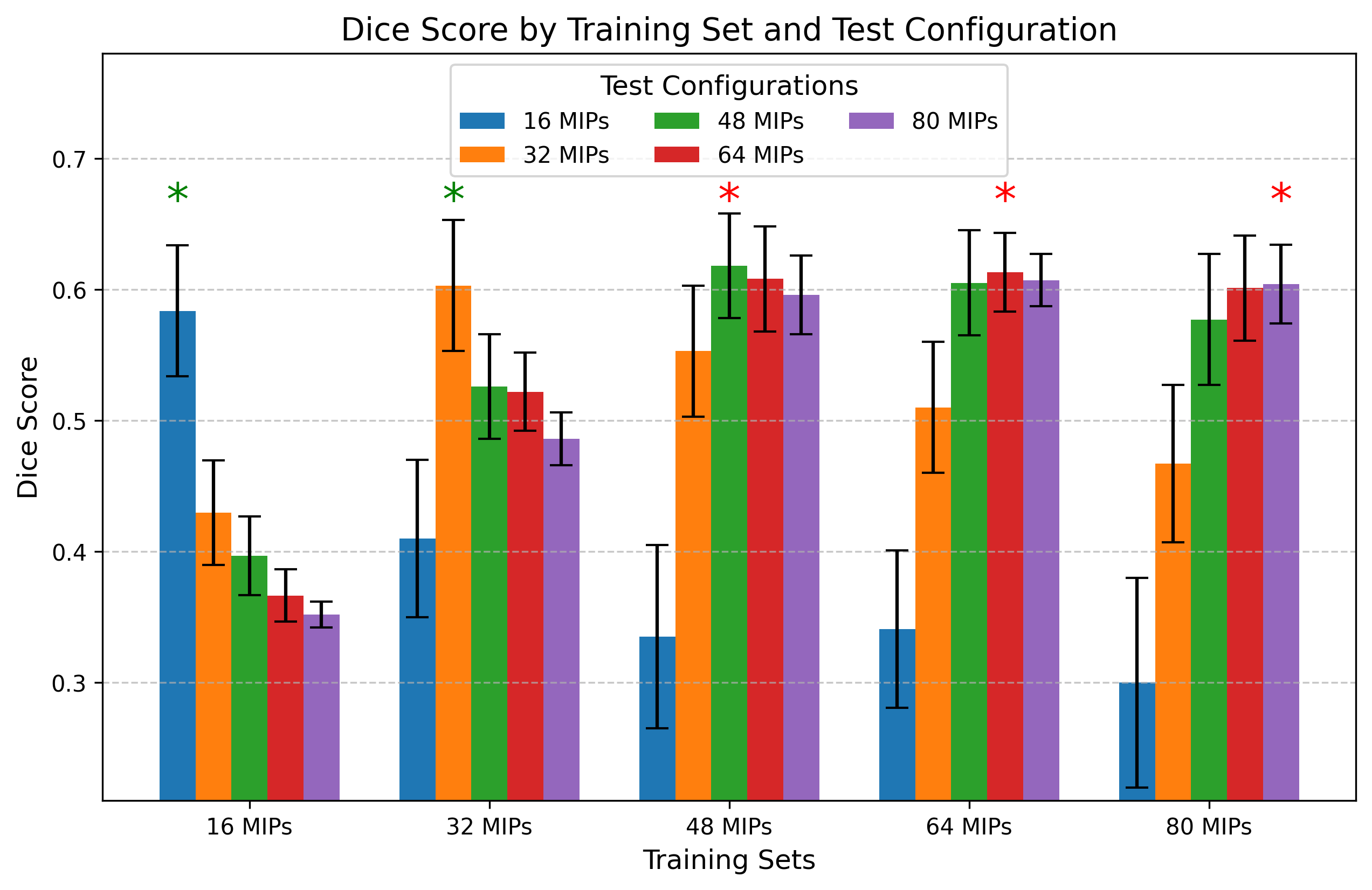}
    \includegraphics[width=5.28cm]{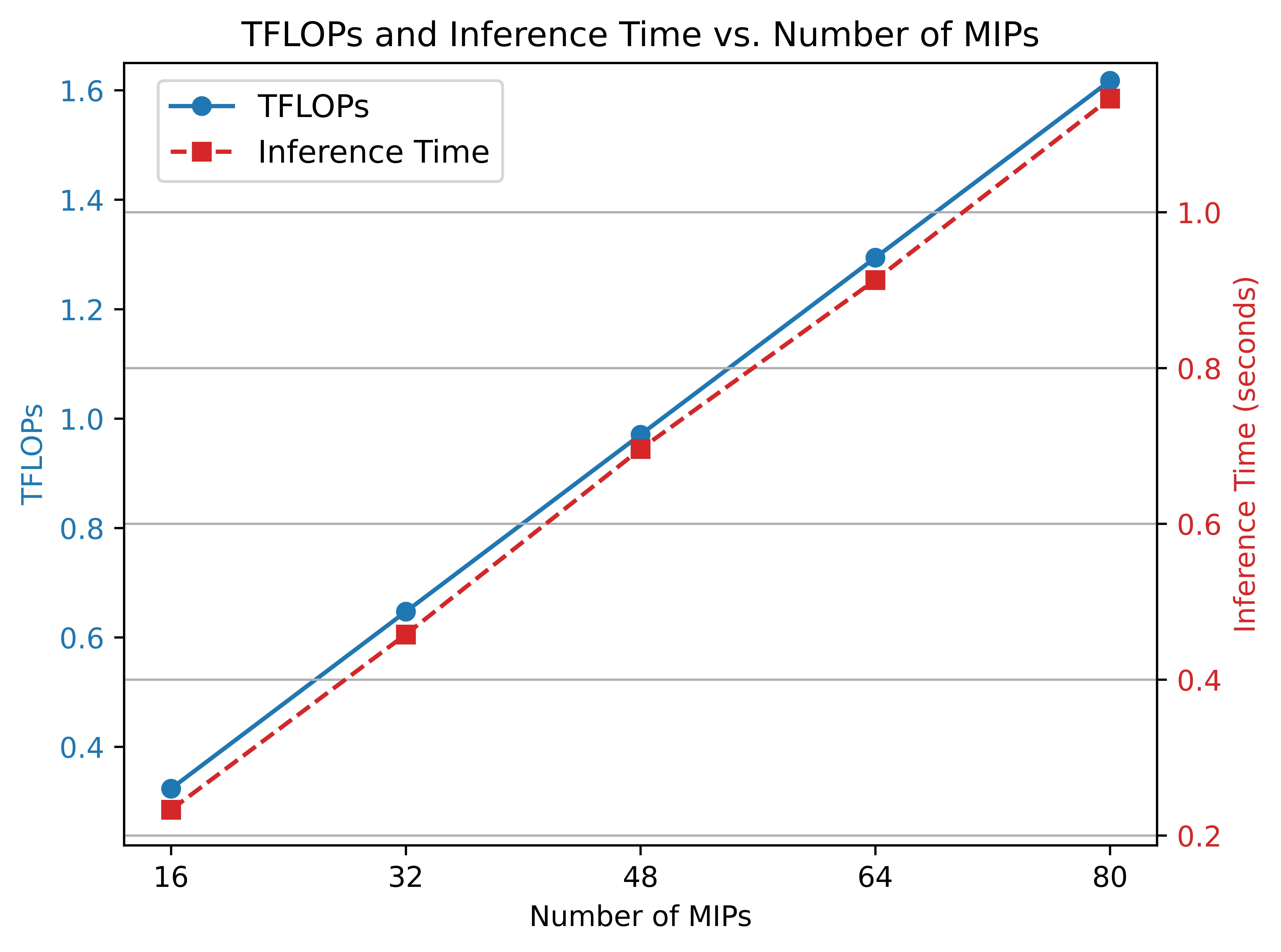}
    \caption{(Left) Dice Scores for models trained/tested on X, Y $\in \{16, 32, 48, 64, 80\}$. (Right) TFLOPs and inference time (seconds) vs.\ number of MIPs. Green (red) stars indicate statistically significant (insignificant) improvement of the highest bar over the second highest, based on a one-sided Wilcoxon signed-rank test.}
    \label{fig:num_of_MIPs_comparison_charts}
\end{figure}

To assess the impact of the number of MIPs on semantic lesion segmentation, we trained models on datasets with different MIP counts and evaluated them on a fixed test set with different MIP counts (see Fig.~\ref{fig:num_of_MIPs_comparison_charts}). This ensured a rigorous analysis while maintaining a fair comparison across different MIP configurations.

\noindent Our results show that 48 MIPs yield the highest and most consistent Dice scores across all train/test combinations (Fig.~\ref{fig:num_of_MIPs_comparison_charts}). Models trained on 48 MIPs performed comparably even when tested on 64 or 80 MIPs, demonstrating both robustness and efficiency. For statistical analysis, a one-sided Wilcoxon signed-rank test was performed by pairing the highest and second-highest test MIPs datasets for each training set, with each pair corresponding to a shared training fold. This approach is appropriate as the test assumes a connection between the pairs; in this case, each pair was trained on the same data, differing only in the number of MIPs used during testing. The analysis revealed statistically significant results for the 16 and 32 MIPs training sets. In contrast, no statistical significance was observed for the training sets with the larger number of MIPs (48, 64, 80).

As shown in the right panel of Fig.~\ref{fig:num_of_MIPs_comparison_charts}, both TFLOPs and average inference time per case increase with the number of MIPs, as expected. Since Dice performance peaks and plateaus at 48 MIPs, we identify this configuration as the optimal trade-off between performance and efficiency.

\section{Conclusions \& Discussion}

We focused and analyzed a radiologist-aligned approach that performs tumor segmentation directly on Multi-Angle MIPs, offering substantial efficiency gains without sacrificing accuracy. Compared to 3D volumetric segmentation, our MIP-based models achieve similar Dice scores (within 1\%) while reducing training time by up to 4.1×, energy per epoch by up to 4.2×, and TFLOPs by over 300×. Classification results further support the approach: using just 16 MIPs, we surpass 3D performance in accuracy and F1-score, with over 10× faster training and approximately 6.65\% energy used. Our analysis shows that 48 MIPs offer the best trade-off between segmentation performance and efficiency. These results suggest MIP-based models are highly efficient and practical for clinical use, especially in resource-constrained settings.

Future work includes incorporating CT into MIPs by mapping corresponding CT voxels from projected PET voxels, and developing models that leverage MIP-specific structural priors. Additionally, \cite{toosi2024segment} proposed MIP segmentation using diffusion models followed by 3D annotation reconstruction via OSEM in PSMA PET scans, a direction we plan to explore on FDG-PET data to enable fair comparisons in the 3D domain. \\

\noindent \textbf{Acknowledgements.} This work was supported by a grant from the Tel Aviv University Center for AI and Data Science (TAD). \\

\noindent \textbf{Disclosure of Interests.} The authors have no competing interests to declare
that are relevant to the content of this article. 

%

\IfFileExists{Paper-0030.bbl}{%

}{%
  \bibliographystyle{splncs04}%
  \bibliography{Paper-0030}%
}

\end{document}